\documentclass[showpacs,preprintnumbers,amsmath,floatfix]{revtex4}
\usepackage{amssymb}

\usepackage{graphicx}
\usepackage{amsmath}
\usepackage{float}
\usepackage{caption}
\usepackage{rotating}
\usepackage[normalem]{ulem}

\begin{document}

\title{Dynamics of EEG Entropy: beyond signal plus noise}
\author{M. Ignaccolo$^{1}$, M. Latka$^{2}$, W. Jernajczyk$^{3}$, P. Grigolini$^{4}$
and B.J. West$^{1,5}$ \\
1) Physics Department, Duke University, Durham, NC\\
2) Institute of Biomedical Engineering, Wroclaw University of Technology,
Wroclaw, Poland \\
3) Department of Clinical Neurophysiology, Institute of Psychiatry and
Neurology, Warsaw, Poland \\
4) Center for Nonlinear Science, University of North Texas, Denton, TX\\
5) Mathematics and Information Science Directorate, Army Research Office}
\date{\today}

\begin{abstract}
EEG time series are analyzed using the diffusion entropy method. The
resulting EEG entropy manifests short-time scaling, asymptotic saturation
and an attenuated alpha-rhythm modulation. These properties are faithfully
modeled by a phenomenological Langevin equation interpreted within a neural
network context. Detrended fluctuation analysis of the EEG data is compared
with diffusion entropy analysis and is found to suppress certain important
properties of the EEG time series. 
\end{abstract}

\maketitle


\section{Introduction}

The mammalian brain generates a small but measurable electrical signal;
first measured in small animals by Caton in 1875 and in people by Berger in
1925. The trace left on a strip chart by this amplified signal was called an
electroencephalograph and the term electroencephalogram (EEG) has
subsequently been used to identify the electrical signal. The power
associated with the EEG signal is distributed over the frequency interval
0.5 to 100 Hz, with most of it concentrated in the interval 1 to 30 Hz. A
typical EEG signal looks like a random time series with contributions from
every part of the spectrum appearing with random phases. In the nearly one
hundred years since the electroencephalogram (EEG) was introduced into
neuroscience there have been a variety of methods used in attempts to
establish a taxonomy of EEG patterns in order to delineate the
correspondence between brain wave patterns and brain activity. The
mathematician Norbert Wiener proposed \textit{generalized harmonic analysis} 
\cite{wiener58} as the mathematical tool necessary to penetrate the
mysterious relations between the EEG time series and the functioning of the
brain. Subsequently, spectral methods have figured prominently in
characterizing the properties of EEG time series. More recently nonlinear
processing techniques, with their implicit dependence on nonlinear dynamics,
chaos and fractals have lead the parade of methodologies hoping to
accomplish this task, see, for example, West \cite{west90} for a brief
review. The progress in relating EEG patterns to brain function has been
slow and the understanding and interpretation of EEG signals remain elusive.
However certain properties of EEG signals have revealed themselves over time.

Over the past half century the single channel EEG time series has been
interpreted as consisting of relatively slow regular variations called 
\textit{signal}, which is the integrated contribution of the neurons in the
vicinity of the channel lead along with the 'coherent' influence of distant
neurons, and the relatively rapid erratic fluctuations called \textit{noise}, 
which is the 'incoherent' contribution of the distal neurons in the brain.
However, the erratic behavior of the EEG time series is so robust that it
persists through all but the most drastic situations including near-lethal
levels of anesthesia, several minutes of asphyxia and the complete surgical
isolation of a slab of cortex \cite{freeman75}. This separation implies that
the signal contains information about the particular neurons associated with
the EEG channels in the brain, whereas the erratic fluctuations are a
property of a channel's environment and does not contain any useful
information. Recent studies have refined this engineering model of signal
plus noise and extracted information from the random fluctuations by
concentrating on what is believed to be the scaling behavior of EEG time
series.

A number of research groups \cite{hwa02,lee02,watters04,stead05,yuan06} have
recently determined that EEG time series $X(t)$ have scaling properties,
with a second moment that increases as a non-trivial power-law, that is, $%
\left\langle X(t)^{2}\right\rangle \varpropto t^{2H}.$ Here the brackets
denote a suitably defined averaging over the data. In random walk models of
classical diffusion the scaling exponent would be given by $H=0.5$, whereas
sub-diffusion processes would have $H<0.5$ and super-diffusion processes
would have $H>0.5$. 
The index $H$ was introduced by Mandelbrot into the study of the long-time
memory of the statistical fluctuations of time series in recognition of
Hurst who first observed such an effect in the yearly river flow of the
Nile. The spectrum $S(f)$ associated with such time series fall in the
category of $1/f$ noise, that is, the Fourier transform of the
autocorrelation function is given by $S(f)\varpropto 1/f^{\beta }$ with
frequency $f$ and power-law index related to the Hurst exponent by $\beta
=2H-1$, consequently the if the index falls within the interval $0\leq \beta
\leq 1$ the underlying process is super-diffusive. However the spectral
index can be greater than one for general complex phenomena such as EEG time
series. In general however the spectral approach is not reliable because the
EEG time series are non-stationary and consequently their direct Fourier
transforms are ill-defined.

The 'signal' parts of the EEG time series are called waves or rhythms. The
nature and scope of these waves have been widely investigated, see Ba\c{s}ar 
\cite{basar06} for a review. The alpha rhythm (7-12 Hz) has been shown to be
typical of awake individuals when the brain is under no stimulation. Ba\c{s}%
ar et al.\cite{basar97} have developed an integrative theory of alpha
oscillations in brain functioning. They hypothesize that there is not one,
but several alpha-wave generators distributed within the brain and note that
the alpha rhythm may act as a nonlinear clock in the manner suggested by
Wiener \cite{wiener58} to serve as a gating function to facilitate the
association mechanisms within the brain.

The method of choice \cite{hwa02,lee02,watters04,stead05,yuan06}, used to
address the issue of non-stationarity in EEG signals, is detrended
fluctuation analysis (DFA) which provides a measure of the standard
deviation of the detrended fluctuations \cite{dfa01}. This method of
processing EEG time series to determine scaling behavior consistently finds
fractal properties, such as those observed in earlier studies. DFA has been
used to quantify the scaling property of EEG dynamics by direct application
to EEG signals \cite{lee02,yuan06} and to EEG increments \cite{hwa02}.
Buiatti et al. \cite{buiatti} use DFA to show that specific task demands can
modify the temporal scale-free dynamics of the ongoing brain activity as
measured by the scaling index. Watters and Martin \cite{watters04}
recognized the two scaling regimes observed in the processing of EEG time
series by a number of investigators and proposed zero-crossings as an
alternative method of analysis in order to focus on the long-time
correlations in the EEG signal. They dichotomize the EEG series according to
the zero crossing time series and apply DFA to the dichotomized EEG. Stead
et al. \cite{stead05} apply DFA to the energy, that is to the square
modulus, of an EEG signal. Finally, the authors of \cite{cai07} analyzed the
histograms for the DFA detrended EEG time series to estimate the associated
probability density function (\textit{pdf}) and its scaling properties.

The observed scaling in EEG time series is not as straightforward as that
observed in other less complex phenomena. Hwa \cite{hwa02}, for example,
finds that the standard deviation of the EEG fluctuations exhibit two
distinct scaling regions, whose variability was analyzed through a moment
technique that could discriminate between normals and those that had
undergone a stroke with up to 90\% accuracy. Robinson \cite{robinson03}, on
the other hand, in his analysis of EEG time series demonstrated the
existence of scaling up to a point in time after which saturation in the
standard deviation occurrs. He attributed this saturation effect to
``dendritic filtering'' and the DFA averaging the influence of
the shape of the spectrum on the time series. Finally, DFA has been used to
study eventual correlation of the alpha rhythm \cite{hansen01,nikulin05}


Various measures other than the standard deviation, spectrum and the
distribution of zero crossings have been introduced into the study of EEG
time series, each one stressing a different physiologic property thought to
be important in representing the brain's dynamics. One recent measure
introduced to quantify the level of coherence in EEG signals is entropy.
However, the entropy of Boltzmann and the information entropy of Shannon
would not be immediately evident in many of these studies. For example,
Inouye et al. \cite{inouye91} employed spectral entropy, as defined by the
Fourier power spectrum, but the fact that EEG time series are not
stationary, in the sense that the autocorrelation function is not simply a
function of the two-time difference, obviates the use of Fourier transforms.
Schl\"{o}gl et al.~\cite{schogl99} measured the information entropy of 16
bit EEG polysomnograhic records and found it to be in the range of 8-11
bits. Patel et al.~\cite{patel99}, using a combination of fMRI and entropy
maximization, where the probability density in the entropy definition is
replaced with a scaled dipole strength, demonstrated that the generators of
alpha rhythm are mainly concentrated over the posterior regions of the
cortex, consistent with the theoretical speculations of others \cite{basar97}%
. Subsequently, wavelet entropy, in which the probability density is
replaced with the relative wavelet energy, was used by Rosso et al. \cite
{rosso01,rosso07} to study the order/disorder dynamics in short duration EEG
signals including evoked response potentials.

In Section II we introduce stochastic differential equations as a way to
model complex phenomena. The \textit{pdf} determined by the stochastic
equations are used to construct the diffusion entropy (DE), which is shown
to be a viable measure of the dynamic mechanisms introduced into such
stochastic models. DE analysis is also used to process EEG time series data
and consequently suggests a form for the stochastic dynamical equations with
which to model the observed EEG properties. Section III introduces the
signal plus noise paradigm of signal processing, which is arguably the basis
for DFA, recently the favored technique in the neuroscience literature for
the analysis of EEG time series. The EEG models developed in Section II are
shown to lead to much different conclusions regarding the scaling of EEG
data depending on whether the DE or DFA methods are used for their analysis.
The coherence of alpha wave is discussed in Section IV to explain why the
alpha-wave modulation of the EEG entropy is attenuated over time. Finally,
we draw some conclusions in Section V.
\section{EEG Models}\label{eegmodels}

Most recently the authors \cite{massi08} have used the DE method \cite{dea}
to characterize the EEG time series dynamic properties. This technique had
been successfully used previously to discriminate between the contributions
of low-frequency waves (signal) and high-frequency fluctuations (noise) in a
number of other phenomena, for example, to determine the seasonal influence
on the daily number of teen births in Texas \cite{bimbini}; the effect of
solar cycles on the statistics of solar flares \cite{flares}; the influence
of solar dynamics on the average global temperature anomalous fluctuations 
\cite{scafetta03}. Herein we use the DE method to provide insight into the
low/high frequency dynamics of EEG time series and compare those results
with processing of the same data using DFA.

\subsection{Ornstien-Uhlenbeck Langevin equation}

In the physics literature there are two strategies for treating stochastic
phenomena, those involving stochastic differential equations for the dynamic
variables and those involving the partial differential equations of
evolution for the probability density in phase space. The former are called
collectively Langevin equations and the latter Fokker-Planck equations. Here
we use a Langevin equation to model the EEG time series because it enables
us to isolate the various physiologic mechanism that might contribute to the
signal. For review consider the Ornstein-Uhlenbeck (OU) Langevin equation
for a linearly dissipative stochastic process

\begin{equation}
\frac{dX(t)}{dt}=-\lambda X(t)+\xi \left( t\right)  \label{langevin1}
\end{equation}
where the random force $\xi \left( t\right) $ is delta correlated in time
with strength $D$ and has Gaussian statistics

\begin{equation}
\left\langle \xi \left( t\right) \xi \left( t+\tau \right) \right\rangle
=2D\delta \left( \tau \right),  \label{langevin2}
\end{equation}
where the symbol $<$$...$$>$ indicates the ensemble average.
Here there are two physical mechanisms; the heat bath modeling the
environment giving rise to the random force and the dissipation modeling the
average energy extracted from the dynamic system by the environment, see for
example, Lindenberg and West \cite{lindenberg}. In a
physical system the fluctuations and dissipation are interdependent through
the Einstein relation

\begin{equation}
\frac{D}{\lambda }=kT  \label{langevin3}
\end{equation}
in order for the system to be thermodynamically closed. Equation (\ref
{langevin3}) is also known as the fluctuation-dissipation relation, which
defines the equilibrium temperature of the heat bath in terms of the ratio
of the strength of the fluctuations to the dissipation rate.

The variance of the dynamic process

\begin{equation}
\sigma ^{2}\left( t\right) \equiv \left\langle X^{2}\left( t\right)
\right\rangle -\left\langle X\left( t\right) \right\rangle ^{2}
\label{langevin4}
\end{equation}
is given here, using the solution to the OU Langevin equation,

\begin{eqnarray}
\sigma ^{2}\left( t\right) &=&\stackrel{t}{\underset{0}{\int }}%
dt_{1}e^{-\lambda \left( t-t_{1}\right) }\stackrel{t}{\underset{0}{\int }}%
dt_{2}e^{-\lambda \left( t-t_{2}\right) }\left\langle \xi \left(
t_{1}\right) \xi \left( t_{2}\right) \right\rangle  \nonumber \\
&=&\frac{D}{\lambda }\left[ 1-e^{-2\lambda t}\right] .  \label{langevin5}
\end{eqnarray}
Consequently, for $t$$\ll 1/\lambda $ the variance increases linearly with
time

\begin{equation}
\underset{t\rightarrow 0}{\lim }\sigma ^{2}\left( t\right) \approx 2Dt
\label{langevin6}
\end{equation}
so that the scaling index is $H$$=$$0.5$. For $t$$\gg 1/\lambda $ the
variance becomes time independent

\begin{equation}
\underset{t\rightarrow \infty }{\lim }\sigma ^{2}\left( t\right) =\frac{D}{%
\lambda },  \label{langevin7}
\end{equation}
with a saturation induced by the dissipation. In the engineering literature
the dissipation is called a filter and its influence on the time series is a
negative feedback reducing the difference between the observed and desired
signal. It is this uncontrolled difference, the noise, that is observed in
the asymptotic saturation level given by the ratio of the strength of the
fluctuations and the dissipation rate.

The solution to the Langevin equation defines a trajectory. An ensemble of
such trajectories, generated by the random force, is used to construct the
histogram of the number of trajectories falling in a specified interval to
estimate the \textit{pdf} $p(x,t).$ The \textit{pdf }can then be used to
calculated the information entropy, a quantity introduced in discrete form
for coding information by Shannon \cite{shannon48} and in continuous form
for studying the problem of noise and messages in electrical filters by
Wiener \cite{wiener48}. We use the latter form here, 
\begin{equation}
S(t)=-\int p(x,t)\log _{2}p(x,t)dx.  \label{entropy}
\end{equation}
Given the Gaussian statistics of the random force in the OU Langevin
equation we know that the statistics of the dynamical variable are also
Gaussian. Substituting a Gaussian distribution with a variance $\sigma
^{2}\left( t\right) $ into (\ref{entropy}) we obtain 
\begin{equation}
S(t)=\log _{2}\left( \sqrt{2\pi e}\sigma (t)\right) .  \label{entropy2}
\end{equation}
Consequently, for $t$$\ll 1/\lambda $ using the approximate variance of Eq.~(%
\ref{langevin6}) the entropy increases as 
\begin{equation}
\underset{t\rightarrow 0}{\lim }S(t)=\frac{1}{2}\log _{2}\left( 4\pi
De\right) +\frac{1}{2}\log _{2}t  \label{entropy4}
\end{equation}
and a linear-log plot yields a straight line of slope $H$$=$$0.5$. At the
other extreme $t$$\gg 1/\lambda $ using the approximate variance (\ref
{langevin7}) the entropy reaches the saturation level 
\begin{equation}
\underset{t\rightarrow \infty }{\lim }S(t)=\frac{1}{2}\log _{2}\left( 
\frac{2\pi De}{\lambda }\right) .  \label{entropy5}
\end{equation}
This use of information entropy is suggestive in that it shares certain
properties with the EEG analysis discussed in the Introduction that being
scaling behavior at early times followed by asymptotic saturation. We
exploit these two mathematical properties of the OU Langenvin equation in
the Section IIC to develop a dynamical model of EEG time series.

\subsection{EEG data analysis}\label{eegdata} 

Each single channel recording of the EEG time series
consists of a sequence of $N+1$ data points, and the difference between
successive data points is denoted by $\xi _{j}$ for $j=1,2,..,N$. For the DE
analysis of the EEG data a set of diffusive variables $X_{k}(t)$ is
constructed from the differenced data points in the following way 
\begin{equation}
X_{k}\left( t\right) =\stackrel{k+t}{\underset{j=k}{\sum }}\xi _{j},\text{ 
}k=1,2,..,N-t+1  \label{rndwlk}
\end{equation}
to obtain $M=N-t+1$ replicas of a diffusive trajectory using overlapping
windows of length $t$. An ensemble of such trajectories, generated by the
EEG time series, is used to construct the histogram from the number of
trajectories falling in a specified interval to estimate the \textit{pdf. }%
Note that this is analogous to what we did in the previous section, the
procedural difference is that here we use data to define the trajectory and
not the solution to a Langevin equation. Another difference is that we do
not know the form of the \textit{pdf} that results from the histogram.
However we can anticipate a class of \textit{pdf}'s based on previous
investigations. For example Schl\"{o}gl \textit{et al}.~\cite{schogl99}
observed a deviation from Gaussian behavior with a time-dependent variance,
but with nearly symmetric empirical distribution functions. We assume here,
but latter establish a theoretical model, a simple analytical form for the 
\textit{pdf} of the diffusion process that satisfies the scaling relation: 
\begin{equation}
p(x,t)=\frac{1}{\sigma \left( t\right) }F\left( \frac{x}{\sigma \left(
t\right) }\right) .  \label{scale}
\end{equation}
Note that a Gaussian diffusion process satisfies (\ref{scale}) with a
time-dependent standard deviation $\sigma (t)$. More generally an alpha-stable 
L\'{e}vy process also scales in this way, in which case $\sigma (t)$
is more general than the standard deviation of the underlying process. The
scaling condition so often anticiapted in EEG time series implies that 
\begin{equation}
\sigma (t)=kt^{\delta },  \label{scale2}
\end{equation}
where $k$ is a constant. Substituting Eqs.~(\ref{scale}) and (\ref{scale2})
into Eq.~(\ref{entropy}) yields 
\begin{equation}
S(t)=-\int F(y)\log _{2}F(y)dy+\log _{2}k+\delta \log _{2}t=C+\delta \log
_{2}t.  \label{scale3}
\end{equation}
This diffusion entropy graphed versus time on log-linear graph paper would
increase linearly with slope $\delta$ with an initial level determined by
the constant $C$. Consequently, the way in which the entropy for a time
series scales is indicative of the scaling behavior of the underlying time
series and consequently of the \textit{pdf}. Note that in a simple diffusive
process this scaling index is equal to the one obtained from calculating the
second moments, that is, $\delta$$=$$H$. However, in general, even when the
data do scale the two power-law indices are not necessarily equal and $\delta$$\neq$$H$.

We now consider EEG signals of twenty awake individuals in the absence of
external stimulations (quiet, closed eyes). EEG signals were recorded using
the 10-20 international recording scheme. For eight individuals only the
channels O1,O2,C3 and C4 were recorded, for the remaining twelve all the
channels are available. To have a consistent database, we restrict our
analysis to the channels O1,O2,C3 and C4, which are the channels
traditionally used in sleep studies. The sampling frequency of all EEG
records is 250Hz, and durations of EEG records vary from 55s to 400s with an
average duration of 128.1s.

\begin{figure}[tbp]
\includegraphics[angle=-90,width=1.0\linewidth]{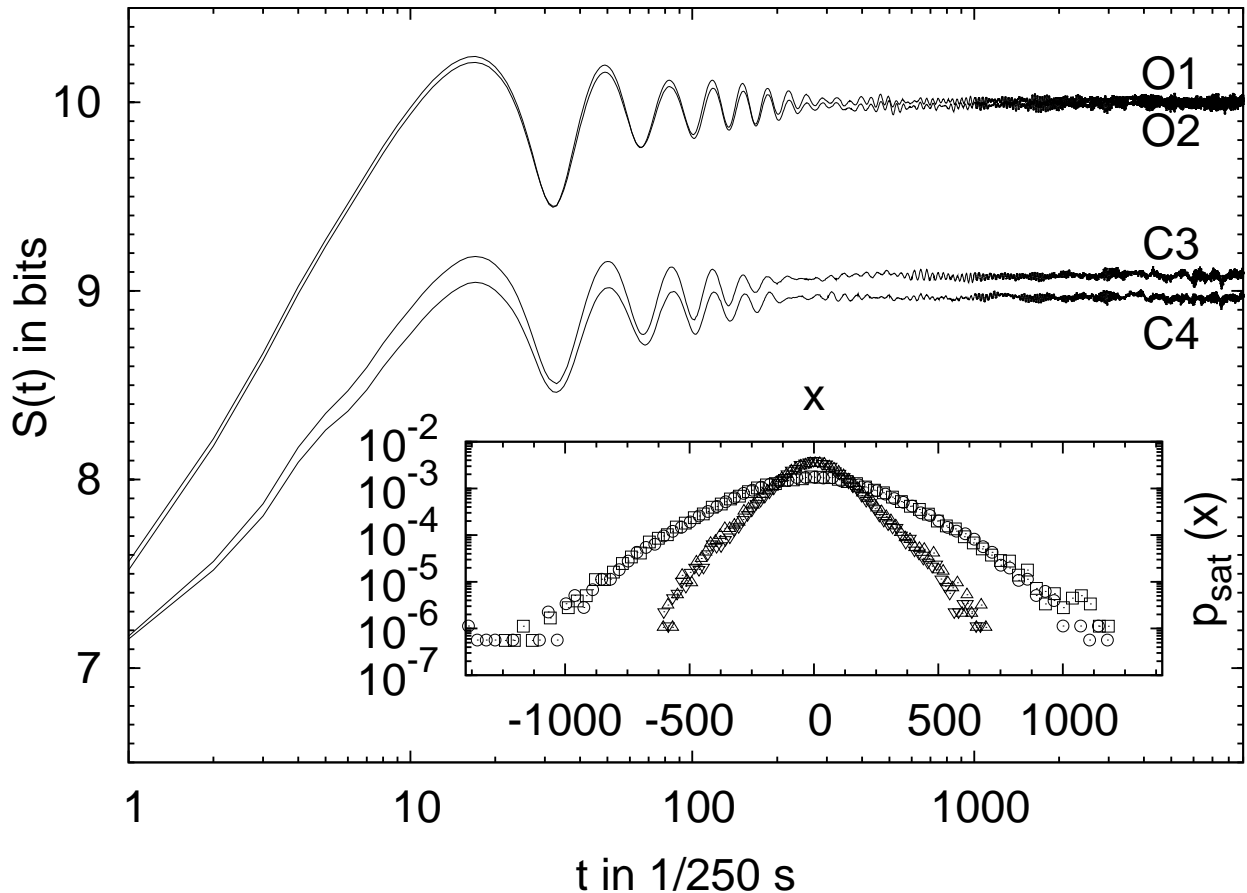}
\caption{The diffusion entropy $S(t)$ calculated using the increments of the
channels O1, O2, C3 and C4 for one of the 20 subjects considered in this
study. The inset depicts the asymptotic pdfs $\text{p}_{\text{sat}}(x)$$=$$p(x,t=2000)$
for each channel: squares (O1), circles (O2), upward triangles (C3), and
downward triangles (C4).}
\label{figure1}
\end{figure}

Fig.~\ref{figure1} shows the DE of the EEG increments for the somnographic
channels O1,O2,C3 and C4 of a single individual. We see how for each channel
the EEG diffusion entropy: 1) reaches a saturation level, 2) has an
``alpha'' ($\sim $7.6 HZ in the case of this individual) modulation which is
attenuated with time, and 3) has a small amplitude residual asymptotic
modulation. The early-time modulation, with variable frequency in the alpha
range and variable amplitude, is observed in the somnographic channels for
all subjects. The saturation effect is present in all channels for all
subjects and it should be pointed out that this saturation is neither a
consequence of the finite length of the time series, nor of the finite
amplitude of the EEG signal. In fact when we randomly rearranged the data
points, thereby destroying any long time correlation in the time series, the
EEG entropy no longer saturates. Consequently, this saturation effect is due
to correlated brain dynamics and is not an artifact of the data processing.
The inset in Fig.~\ref{figure1} depicts the \textit{pdf}s $p_{\text{sat}}(x)$%
, after the entropy saturation is attained. These distributions have
approximately exponential tails. 
\begin{figure}[tbp]
\includegraphics[angle=-90,width=1.0\linewidth]{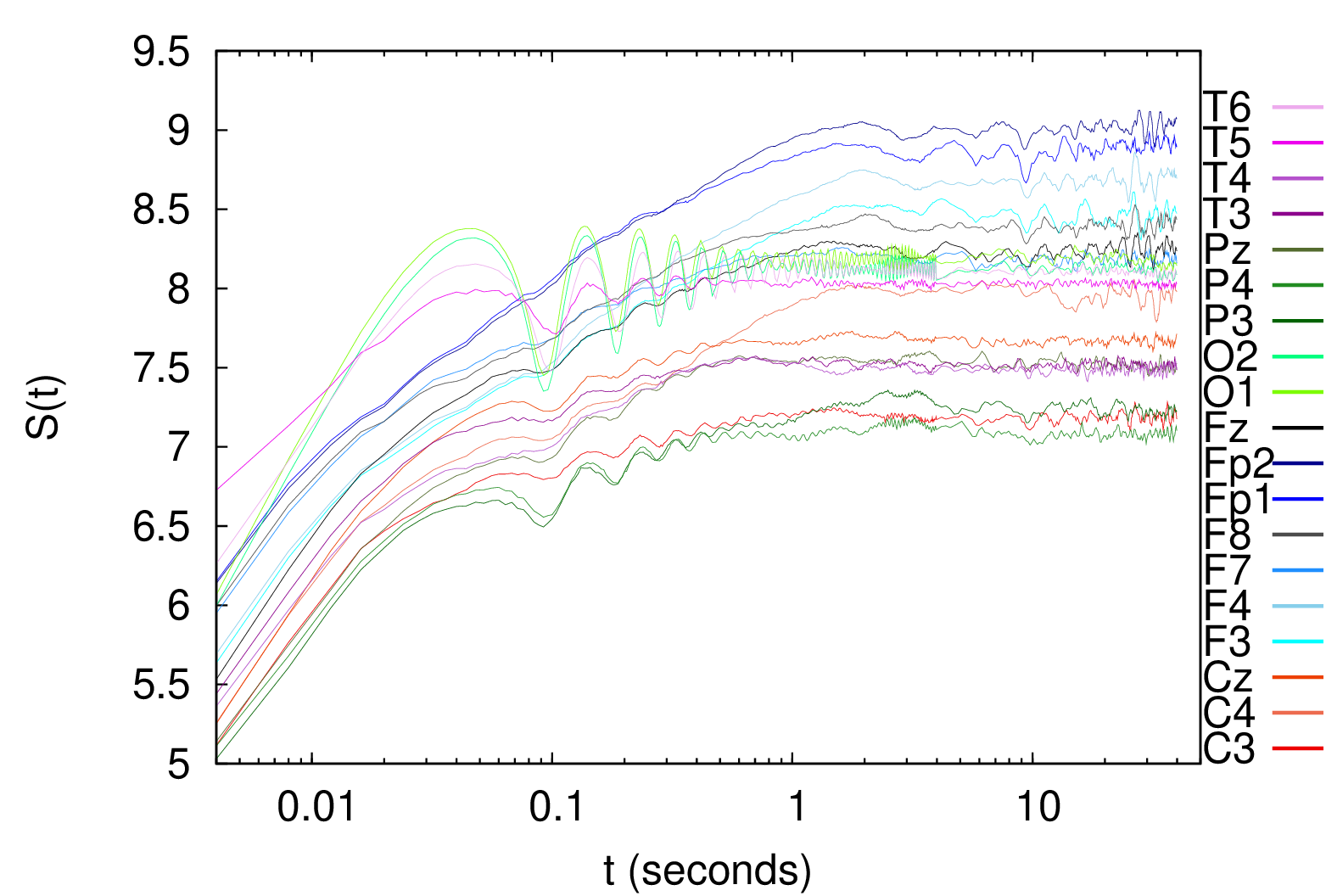}
\caption{The diffusion entropy $S(t)$ of the EEG increments of all the 19
channels for one of the 20 subjects considered in this study.}
\label{figure3}
\end{figure}
Fig. \ref{figure3} depicts the diffusion entropy for the nineteen channels
of a representative individual. In this figure the somnographic channels
have strong alpha rhythms, but the other channels do not. However it is
evident that regardless of the alpha wave content of the EEG time series
each and every channel saturates.

\subsection{EEG Langevin equation}\label{EEGlang} 

The simplest dynamic model, which includes all the
properties identified in Fig.~\ref{figure1}, these being, fluctuations,
modulation and dissipation, has the form of a Langevin equation. We assume a
dissipative linear dynamic process dynamic process $X(t)$, i.e., an OU
process, with a periodic driver having a random amplitude and frequency and
an additive random force $\eta \left( t\right) $ which is a delta correlated
Gaussian process of strength $D$: 
\begin{equation}
\frac{dX(t)}{dt}=-\lambda X(t)+\eta \left( t\right) +\underset{j=0}{\sum }%
A_{j}\chi \left[ I_{j,s}\right] \sin \left[ 2\pi f_{j}t\right] 
\label{sinusoidal}
\end{equation}
The coefficient $\lambda $ is positive definite and defines a negative
feedback, $\chi \left[ I_{j,s}\right] =1$ when the argument of $\chi \left[
{}\right] $ is the time interval $I_{j,s}=[jt_{s},(j+1)t_{s}]$ and is zero
otherwise, and $t_{s}$ is the 'stability' time after which a new constant
frequency $f_{j}$ and a new constant amplitude $A_{j}$ are selected.

The values of the frequencies $f_{j}$ and amplitudes $A_{j}$ are empirical
and determined in the following way. First, we evaluate the spectral density
in the time-frequency domain of time series of EEG increments with a time
resolution $t_{s}$ and a frequency resolution $\Delta f$ by means of a
Windowed Fourier Transform. The theoretical spectral density is estimated
from the spectrogram, which is the spectrum of a time series for a given
time resolution, but which changes as a function of time. Therefore there is
no one spectrum to characterize the process as we sweep through the
non-stationary EEG time series, see the discussion on spectrograms in Ref.~\cite{mallat}. 
The spectral density, or spectrogram, is a three-dimensional
plot of the spectrum of the EEG increments $\xi _{j}$ as it changes over
time. Then, for each time interval of duration $t_{s}$ we consider the range
of frequencies of the alpha waves, 7-12 Hz, and find which frequency has the
maximum amplitude in the spectrogram. This procedure defines the frequency
and the amplitude of the time interval considered.

\begin{figure}[tbp]
\includegraphics[angle=0,width=1.0\linewidth,height=9cm]{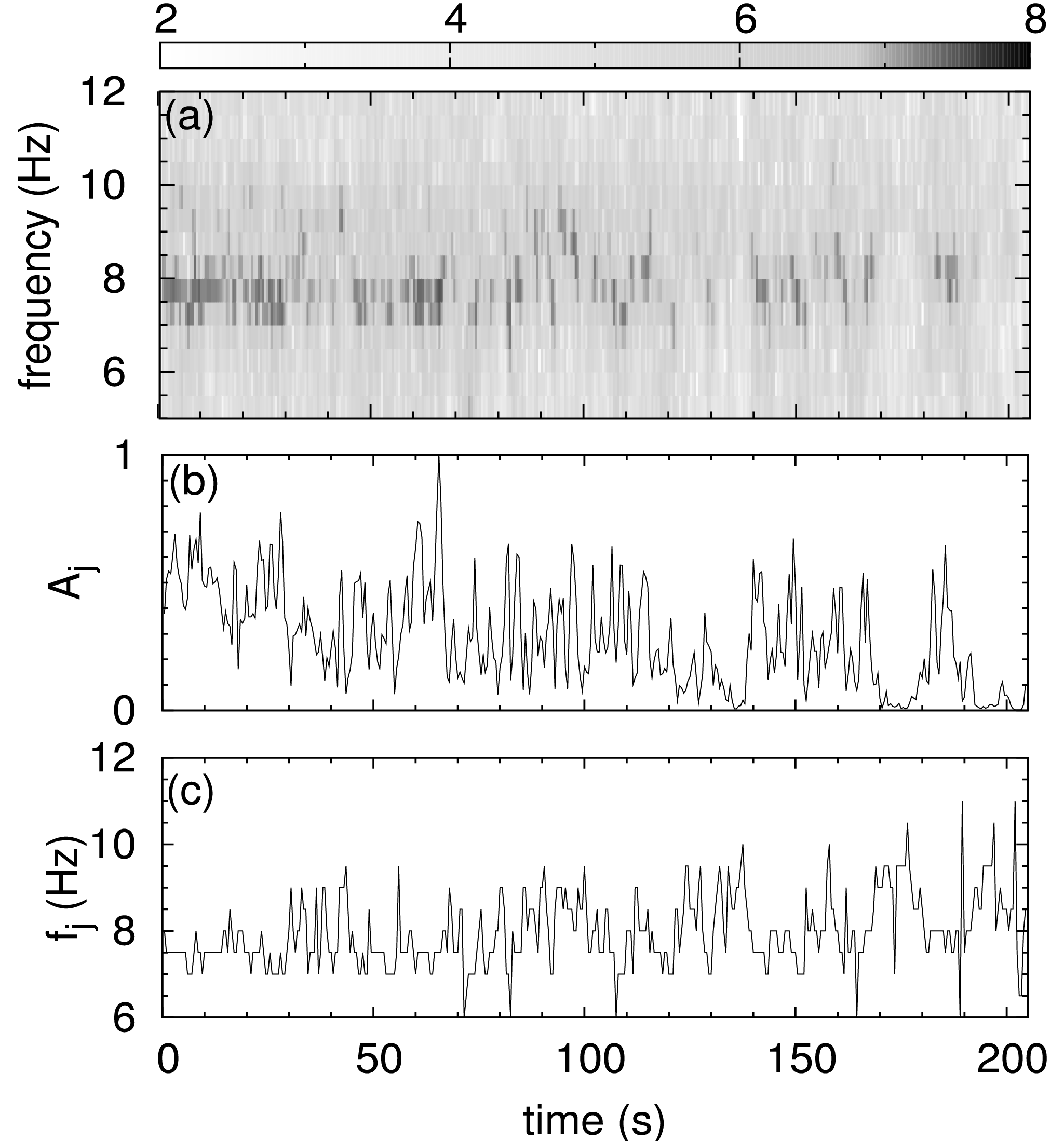}
\caption{(a) Spectrogram of the increments of channel O1. We plot the
base-10 logarithm of the spectral density. The time resolution is $t_{s}$$=$%
0.5 s, the frequency resolution is $\Delta f$$=$0.5 Hz. (b) Sequence of the
maxima of the spectrogram amplitude (normalized so that the maximum
amplitude is 1). This sequence of coefficients $A_{j}$ is used in Eq.~( \ref
{sinusoidal}). (c) Sequence of the frequencies corresponding to the
amplitude maxima of the spectrogram. This sequence of coefficient $f_{j}$ is
used in Eq.~(\ref{sinusoidal}).}
\label{figure4}
\end{figure}

Panel (a) of Fig.~\ref{figure4} shows the spectrogram relative to the
increments $\xi _{j}$ of the channel O1 for the same subject as in Fig.~\ref
{figure1}. Panels (b) and (c) of Fig.~\ref{figure4} show respectively the
sequence of amplitudes $A_{j}$ (normalized to a maximum amplitude of 1) and
of frequencies $f_{j}$ calculated using the procedure described above.
Without an \textit{a priori} knowledge of the typical duration of an alpha
wave packet, we set the stability time $t_{s}$ of Eq.~(\ref{sinusoidal})
equal to 0.5s. A time resolution of 0.5s and a frequency resolution of
approximately 0.5Hz in the spectrogram represent a reasonable time-frequency
localization for our purposes. 

\begin{figure}[h]
\includegraphics[angle=-90,width=1.0\linewidth]{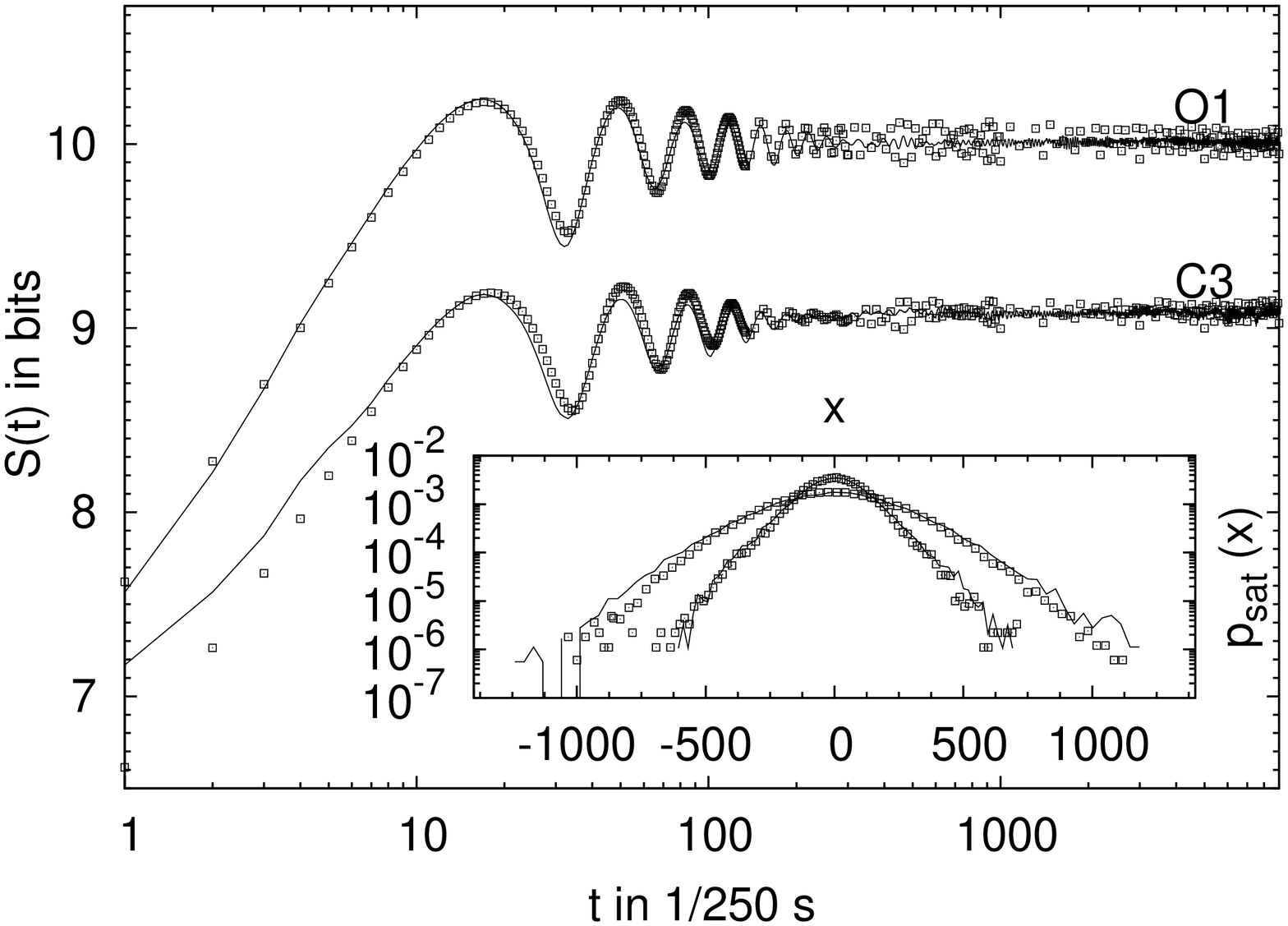}
\caption{Comparison between the diffusion entropy of the increments, solid
lines, of channel O1 and C3, and diffusion entropy of the increments,
points, of the variable $X$ of Eq.~(\ref{sinusoidal}). The parameters used
in Eq.~(\ref{sinusoidal}) are $\lambda $$=$0.055 D$=$40 for O1 and $\lambda $%
$=$0.055 D$=$20 for C3. Inset show the comparison between the asymptotic pdfs 
$\text{p}_{\text{sat}}(X)$$=$$p(X,t=2000)$: channels O1 and C3,
solid lines, variable $X$ of Eq.~(\ref{sinusoidal}), squares.}
\label{figure5}
\end{figure}

When the modulation is present~Eq.(\ref{sinusoidal}) is numerically
integrated, and the increments of the dynamic variable $X$ are processed
using the DE algorithm. In Fig.~\ref{figure5}, we compare the EEG entropy
obtained using the integrated solutions of Eq.~(\ref{sinusoidal}) with that
of the channels O1 and C3, already shown in Fig.~\ref{figure1}. It is evident
that the entropy constructed from the solution to Eq.(\ref{sinusoidal})
captures the qualitative and many of the quantitative features of the DE of
the EEG increments. Moreover, the asymptotic \textit{pdf}s recorded in the
inset also agree with the empirical \textit{pdf}s depicted in Fig.~\ref
{figure1}. 
In Table~1 we average the phenomenological parameters $\lambda $ and $D$ for
the somnographic channels for the twenty subjects in this study. 
\begin{table}[h]
\caption{The average values (avg.) and the standard deviations (s.d.) of the
parameters $\lambda $ and $D$ of Eq.~(\ref{sinusoidal}) for all 20 subjects
in this study.}
\label{table1}
\begin{tabular}{|c|c|c|}
\hline
EEG channel & $\lambda$ (avg.$\pm$s.d.) & D (avg.$\pm$s.d.) \\ \hline
O1 & 0.0461$\pm$0.0187 & 16.37$\pm$6.88 \\ \hline
O2 & 0.0497$\pm$0.0182 & 16.35$\pm$6.72 \\ \hline
C3 & 0.0362$\pm$0.0186 & 10.19$\pm$3.90 \\ \hline
C4 & 0.0393$\pm$0.0200 & 10.60$\pm$3.72 \\ \hline
\end{tabular}
\end{table}
Note that both the strength of the fluctuations and the dissipation rates
change between the O1, O2 values and the C3, C4 values. This suggests that
the channel environment changes in a statisitically significant way from one
region of the brain to another, with the O1, O2 channels being noisier and
more dissipative.

\section{Diffusion Entropy analysis versus DFA}

In Sec.~\ref{eegmodels}, we showed how the properties of the EEG records can
be modeled using an OU Langevin equation: The solution to Eq.~(\ref
{sinusoidal}) if the EEG record has an alpha rhythm (the generalization to a
different rhythm or a sum of two or more rhythms is straightforward), and
the solution to Eq.~(\ref{langevin1}) if no rhythm is present. As stated in
the Introduction, a main tenet of the traditional EEG analysis is the
decomposition of an EEG record $X_{j}$ into the sum of two orthogonal
components: 
\begin{equation}
X_{j}=S_{j}+N_{j}\;\;\text{AND}\;\;\text{Cov}(S,N)=0  \label{sigplusnoise}
\end{equation}
where $S_{j}$ is the time varying mean or \textit{signal} (rhythms), $N_{j}$
is the \textit{noise} or random component and Cov$(S,N)$ is the covariance
of the two. DFA was introduced \cite{dfa01} as a tool to measure the scaling
of the variance of the noise component of a time series $N_{j}$ without 
\textit{a priori} knowledge of the signal component of the time series $%
S_{j}$. For this reason, as well as the fact that it apparently works for
time series that scale, DFA has been widely used in the analysis of EEG time
series.

\begin{figure}[tbp]
\includegraphics[angle=0,width=1.0\linewidth,height=9cm]{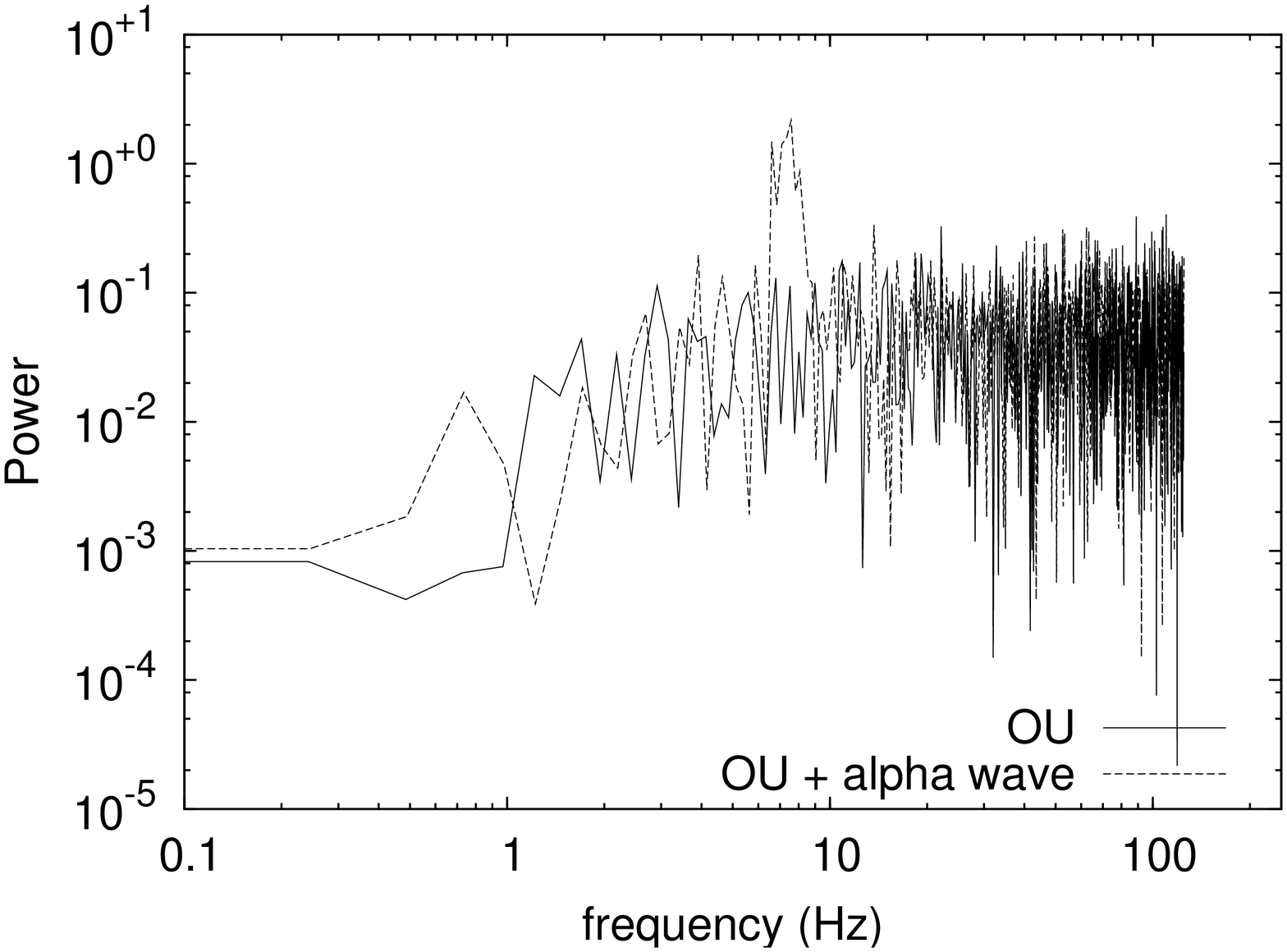}
\caption{ The solide line depicts the power spectrum of the increments of
the variable $X(t)$ of Eq.~(\ref{langsol}) ($\lambda$$=$0.055 and $D$$=$800), the dashed line depicts the power spectrum of the increments of the
variable $X(t)$ of Eq.~(\ref{langpersol}) ($\lambda $$=$0.055, $D$$=$800 and
alpha wave coefficients of the channel O1 of Fig~.\ref{figure1}).}
\label{power}
\end{figure}

However, the signal plus noise decomposition of Eq.~(\ref{sigplusnoise}) is
not applicable to the OU processes or the driven OU processes of Eqs.~(\ref
{langevin1}) and (\ref{sinusoidal}), respectively. The solution to Eq.~(\ref
{langevin1}) is 
\begin{equation}
X(t)=e^{-\lambda t}\int\limits_{0}^{t}\eta (t^{\prime })e^{\lambda t^{\prime
}}dt^{\prime }  \label{langsol}
\end{equation}
where we assumed, without loss of generality, that $X(0)$$=$$0$. In this
case we have no \textit{signal} component. Moreover, Eq.~(\ref{langsol})
states that the present value of $X(t)$ depends on all the previous history.
In particular, the autocorrelation function $\Phi _{\xi }(t)$ of the
increments $\xi (t)$$=$$d/dt$$X(t)$ is negative with an exponential decay: 
\begin{equation}
\Phi_{\xi }(t)=\frac{d}{dt^{2}}\sigma ^{2}(t)=\delta (t)-4D\lambda \exp
\left( {-2\lambda t}\right),  \label{autocorr}
\end{equation}
where $\delta (t)$ is the Dirac delta function. A consequence of Eq.~(\ref
{autocorr}) is that the power spectrum of the increment time series $\xi (t)$
does not satisfy the relation $S(f)\varpropto 1/f^{\beta }$. Applying the
Wiener$-$Khintchine theorem to Eq.~(\ref{autocorr}), we obtain 
\begin{equation}
S(f)=\frac{1}{2}\left[ 1-\frac{4\lambda ^{2}}{4\lambda ^{2}+4\pi ^{2}f^{2}}%
\right]   \label{powerlang}
\end{equation}
and consequently the spectrum is zero at low frequencies and increases
quadratically to a constant value at high frequencies. But let us consider
the more general case, using the solution of Eq.~(\ref{sinusoidal}),
assuming $X(0)$$=$$0$, 
\begin{equation}
X(t)=e^{-\lambda t}\int\limits_{0}^{t}\left\{ \eta (t^{\prime })+\underset{%
j=0}{\sum }A_{j}\chi \left[ I_{j,s}\right] \sin \left( 2\pi f_{j}t^{\prime
}\right) \right\} e^{\lambda t^{\prime }}dt^{\prime }.  \label{langpersol}
\end{equation}
If we identify the alpha-wave component, the second term under the integral,
as the \textit{signal} then it is \textit{not} possible to separate $X(t)$
according to Eq.~(\ref{sigplusnoise}). In particular the covariance Cov$(S,N)$ 
does not vanish because the present value of $X(t)$ depends on its
previous values. As for the covariance function of the increments in this
case, the behavior of Eq.~(\ref{autocorr}) is periodically modulated, while
the power spectrum shows a peak in the alpha range as shown in Fig.~\ref
{power}. 

For these reasons, in this section, we compare the results of DFA and DE
when applied to EEG records, and to records (via numerical integration) of
the OU processes of Eqs.~(\ref{langevin1}) and (\ref{sinusoidal}). We set the lenght of these records to be 
50,000 to match the typical length of
the EEG records examined herein: 50,000 data points is equivalent to 200
seconds of data with a 250Hz sampling frequency.

First, we briefly describe the DFA algorithm. Given a time series $X_{k}$,
the zero-averaged time series is aggregated 
\begin{equation}
Y_{j}=\sum\limits_{k=1}^{k=j}\left[ X_{k}-X_{avg}\right]   \label{dfastep1}
\end{equation}
where $X_{avg}$ is the average of the time series $X_{k}$. The integrated
signal is divided into windows of size $t$. Here we use overlapping windows
adopting the same procedure as the DE algorithm (Sec.~\ref{eegdata}), while
the original algorithm uses non-overlapping windows \cite{dfa01}. For each
window, a least-squares fit is computed with a polynomial of order $n$$\geq $%
1. This fitting procedure eliminates the local trend: the \textit{signal} in
that porticular window. Finally the local
trend is subtracted from the integrated time series and the standard
deviation of the residuals $\tilde{Y}_{j}$ calculated: 
\begin{equation}
F(t)=\sqrt{\frac{1}{\text{t}}\sum\limits_{j=1}^{\text{t}}\tilde{Y}_{j}^{2}}.
\label{dfastep2}
\end{equation}
These steps are repeated for increasing values of the window size $t$. The
scaling condition for the standard deviation implies 
\begin{equation}
F(t)\propto t^{\alpha }\Leftrightarrow \log _{2}F(t)\propto \alpha \log _{2}t
\label{dfastep3}
\end{equation}
where the scaling parameter is in the interval $0\leq \alpha \leq 1$.

\begin{figure}[tbp]
\includegraphics[angle=-90,width=1.0\linewidth]{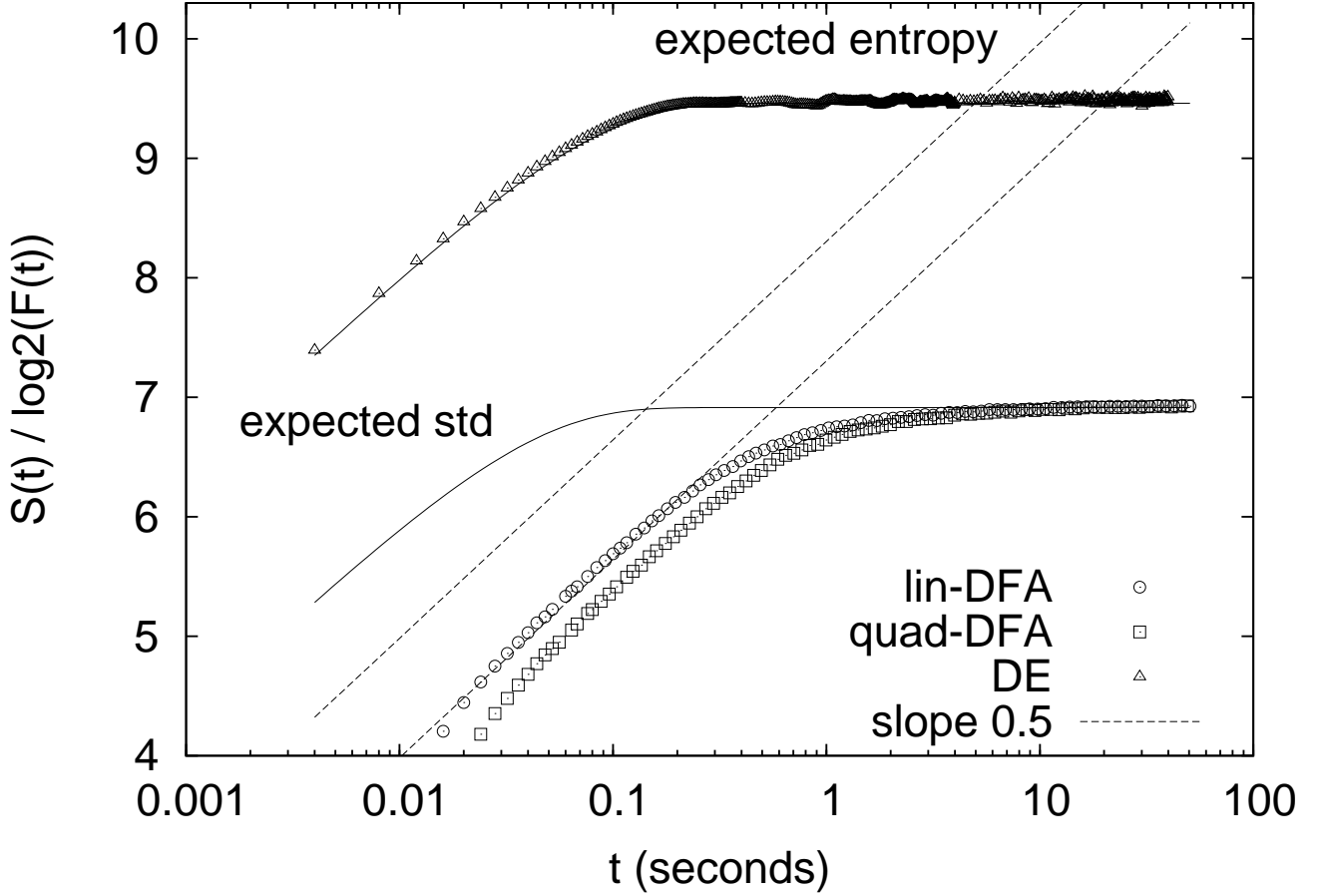}
\caption{DE and DFA for the increments of the variable $X(t)$ of Eq.~(\ref
{langevin1}) with $\lambda $$=$0.055 and $D$$=$800. The triangles indicates
the values of the entropy $S(t)$, while circles (linear detrending) and
squares (quadratic detrending) are the values of base-2 logarithm of the
variance $F(t)$. The solid lines are the expected values for $S(t)$ and $%
\log _{2}(F(t))$. The dashed line indicates a logarithmic increase with a
slope 0.5.}
\label{figure6}
\end{figure}

In Fig.~\ref{figure6}, we plot the results of DE and DFA when applied to the
time series of the increments of the variable $X(t)$ of Eq.~(\ref{langevin1}) 
with $\lambda$$=$$0.055$ and $D$$=$$800$. The DE (triangles) agrees \cite{overlaperror} 
with the theoretical prediction of Eq.~(\ref{entropy2}) (solid line).
Moreover, the DE results show that the approximation given by Eq.~(\ref
{entropy4}) for the case $\lambda$$=$$0.055$ is valid up $\sim$0.02$-$0.03s
(note that the factor $1$/$\lambda $ must be divided by the value of
sampling frequency which is 250Hz). DFA results (circles for a linear
detrending and squares for a quadratic detrending) show hints of a
saturation regime more than two decades later than it actually occurs: $\sim
10s$ instead of $\sim 0.1s$. As for the expected initial regime $F(t)\propto 
$$\sqrt{t}$ (Eq.~(\ref{langevin6})), one can linearly fit the DFA curves in
different ranges for times $t\lesssim 0.4s$ (before the strong ``bending''
occurs in the DFA). The results depend on the particular fitting range used:
e.g. in the range $0.04s\leq t\leq 0.4s$ the resulting slope (for the DFA
curve obtained with a linear detrending) is $\simeq 0.44$. 

\begin{figure}[tbp]
\includegraphics[angle=-90,width=1.0\linewidth]{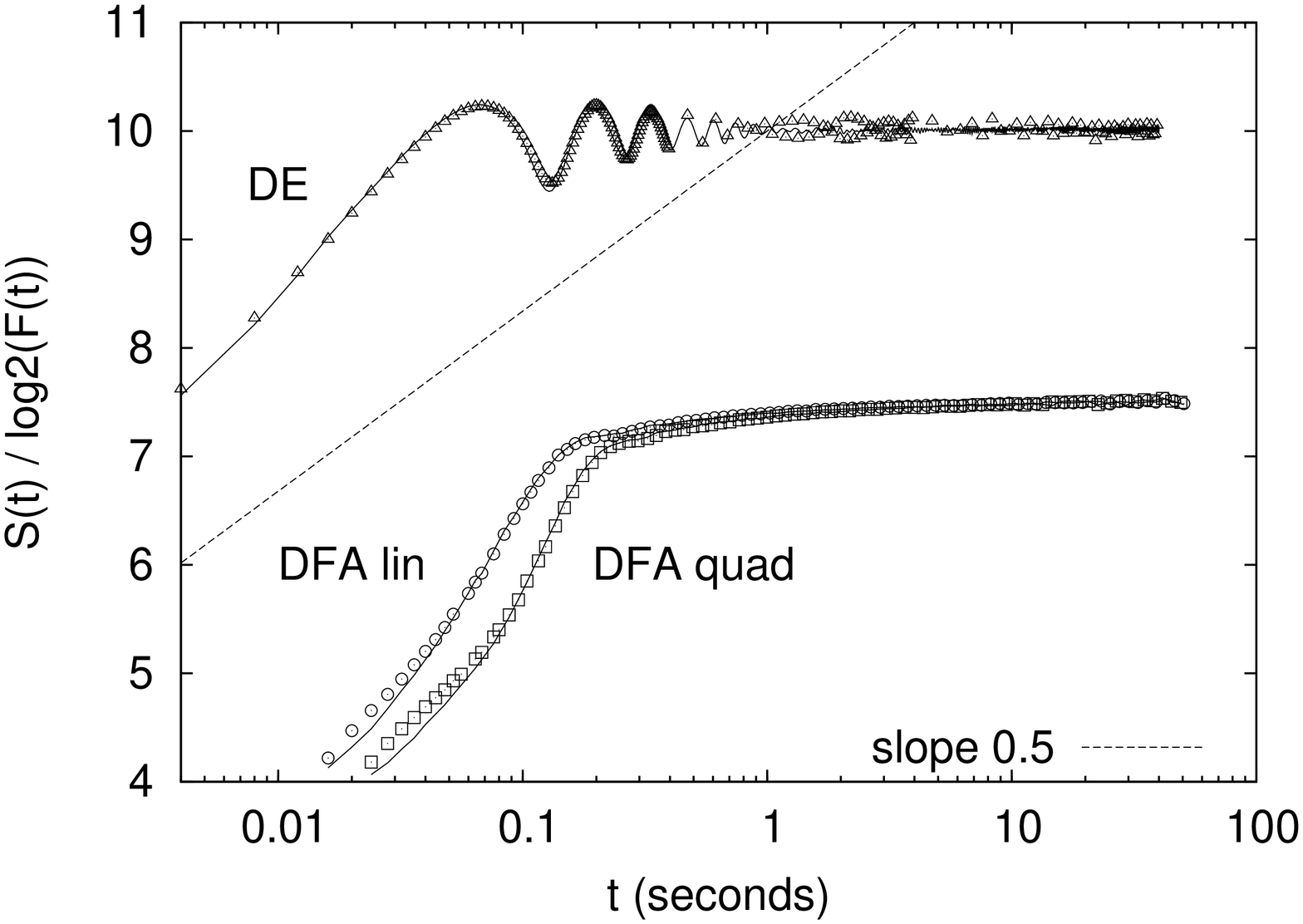}
\caption{DE method and DFA for the increments of a EEG record (channel O1)
and the increments of its best approximation via the model of Eq.~(\ref
{sinusoidal}) (Section \ref{EEGlang}). Solid lines indicate the results of
DE and DFA relative to the EEG record. Trianlges indicate the results of DE
calculation relative to the Ornstein-Uhlenbeck process of Eq.~(\ref
{sinusoidal}) approximating the EEG record, while circles (linear
detrending) and squares (quadratic deternding) indicate those of DFA.
Finally, the dashed line indicates a logaritmic increase with a slope of
0.5. }
\label{figure7}
\end{figure}

Fig.~\ref{figure7} shows the results of the DE method and DFA for the
increments of a EEG record (channel O1) and the increments of its best
approximation via the model of Eq.~(\ref{sinusoidal}) (Section \ref{EEGlang}%
). We notice how the results, for both DE ad DFA, relative to the OU process
of Eq.~(\ref{sinusoidal}) (triangles for DE and circles and squares for DFA)
reproduce those relative to the EEG record (solid line). We notice how the
presence of the alpha rhythm results in an initial ($t<0.04s$) slope of the
diffusion entropy $S(t)$ which is larger than 0.5 as expected with no alpha
rhythm Eq.~(\ref{entropy4})): for times smaller than the typical period of
the alpha rhythm, the alpha rhythm is ``equivalent'' to a trend which
produce an additional entropy increase (for detailed discussion of this
effect see \cite{bimbini}, for example. As for the results of DFA, we see
how the modulation due to the alpha-wave packets has been eliminated and
instead two ``slopes'' are observed for $t\lesssim 0.2s$. As in Fig.~(\ref
{figure6}) DFA approaches a saturation regime two decades later than what
expected from the model of Eq.~(\ref{sinusoidal}) and correctly detected by
the DE. For the DFA curve obtained with a quadratic detrending, we have:
slope 0.65 for $0.01s<t<0.06s$, slope 1.2 for $0.08s<t<0.2s$, and slope 0.1
for $0.3s<t<1.6s$. These results indicate that the DFA is not able to
accurately ``detrend'' the periodic component (alpha wave) and the observed
linear regimes are not actual scaling regimes. 
In Figs.~\ref{figure6} and \ref{figure7}, we have applied the DE and the DFA
to the increments of the variable $X(t)$ (Eq.~(\ref{langevin1}) for Fig.~\ref
{figure6} and Eq.~(\ref{sinusoidal}) for Fig.~\ref{figure7}) or to the EEG
increments $\xi _{j}$. Some investigators \cite{hwa02} use DFA to analyze
the resting (closed eye) EEG of normal subjects and of subjects with acute
ischemic stroke. They report the presence in virtually all channel and for
all subjects of a ``double'' scaling regime in the time ranges corresponding
to the second and third linear regimes of Fig.~\ref{figure6}. The reported 
\cite{hwa02} ranges of values for the scaling parameters are compatible with
the one calculated for Fig.~\ref{figure6}. However the analysis presented in
this manuscript suggest that these linear regimes observed in the DFA are
not genuine scaling regimes but are the result of the EEG dynamics not
satisfying the signal plus noise decomposition of Eq.~(\ref{sigplusnoise})
thereby compromising the capability of the DFA to detrend the alpha-wave
component.

Most studies in the literature, however, report results of the application
of DFA to the EEG time series itself. Thus, we too apply DE and DFA to the
variable X(t) of Eqs.~(\ref{langevin1}) and (\ref{sinusoidal}), and to the
EEG itself as well. 
\begin{figure}[tbp]
\includegraphics[angle=-90,width=1.0\linewidth]{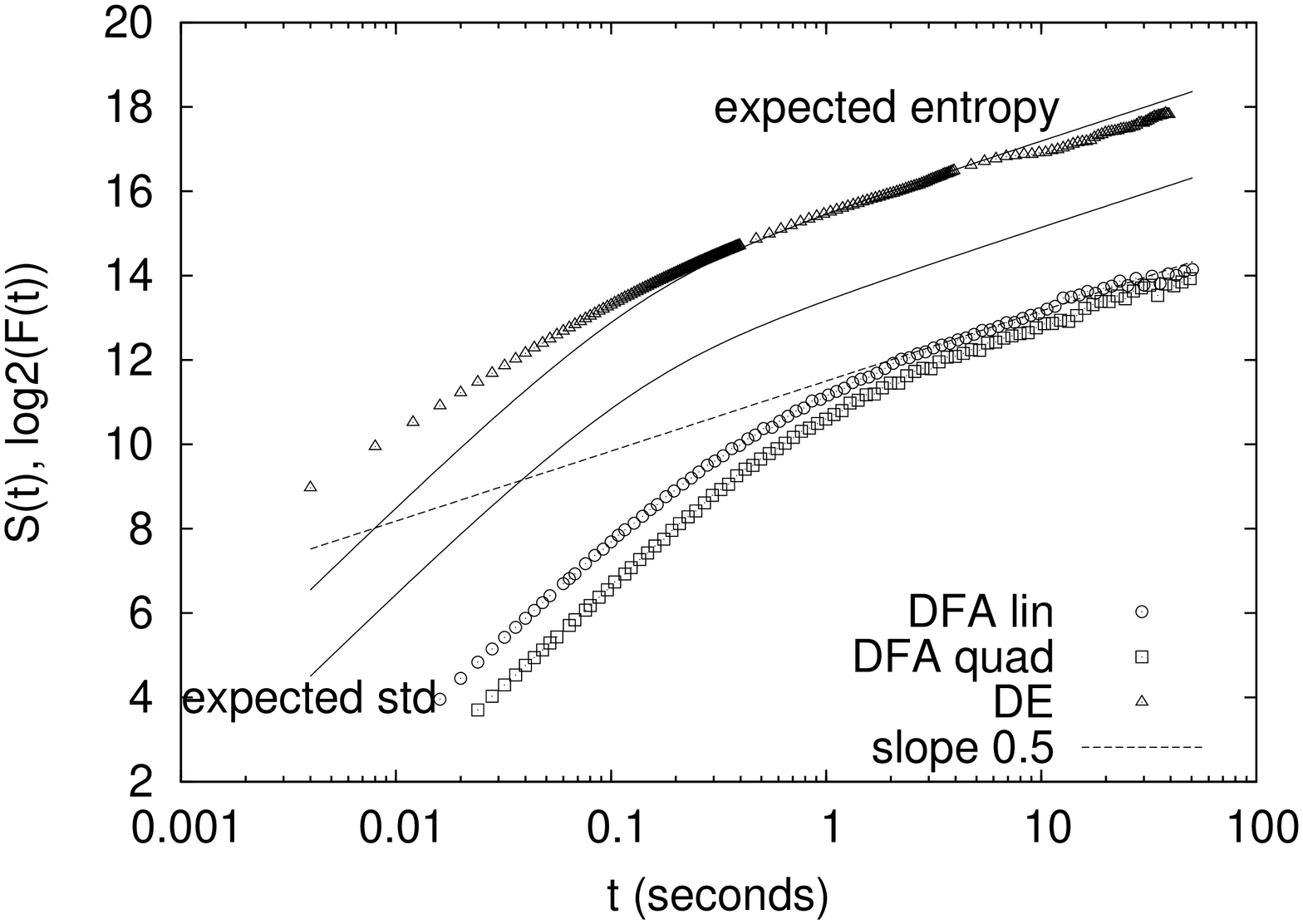}
\caption{DE and DFA for the variable $X(t)$ of Eq.~(\ref{langevin1}) with $%
\lambda $$=$0.055 and $D$$=$800. The triangles indicates the values of the
entropy $S(t)$, while circles (linear detrending) and squares (quadratic
detrending) are the values of base-2 logarithm of the variance $F(t)$. The
solid lines are the expected values for $S(t)$ and $\log _{2}(F(t))$. The
dashed line indicates a logarithmic increase with a slope 0.5.}
\label{figure8}
\end{figure}

Fig.~\ref{figure8} show the application of DE and DFA to the variable $X(t)$
satisfying Eq.~(\ref{langevin1}). The solid lines represent the expected
values for the standard deviation. These are obtained by integrating the
variable $X(t)$ 
\begin{equation}
Y(t)=\int\limits_{0}^{t}X(t^{\prime })dt^{\prime
}=\int\limits_{0}^{t}dt^{\prime }e^{-\lambda t^{\prime
}}\int\limits_{0}^{t^{\prime }}\eta (t^{^{\prime \prime }})e^{\lambda
t^{^{\prime \prime }}}  \label{langint}
\end{equation}
and calculating the variance $\sigma _{Y}^{2}(t)$: 
\begin{equation}
\sigma _{Y}^{2}(t)=\frac{D}{\lambda ^{3}}\left[ 2\lambda t+4e^{-\lambda
t}-e^{-2\lambda t}-3\right] .  \label{varY}
\end{equation}
Using Eq.~(\ref{varY}), we obtain for times $t$$\ll $$1/\lambda $ 
\begin{equation}
\underset{t\rightarrow 0}{\lim }\sigma _{Y}^{2}(t)=\frac{2}{3}Dt^{3},
\label{varY1}
\end{equation}
while for $t$$\gg $$1/\lambda $ 
\begin{equation}
\underset{t\rightarrow \infty }{\lim }\sigma _{Y}^{2}(t)=\frac{2D}{\lambda
^{2}}t-\frac{3D}{\lambda ^{3}}.  \label{varY2}
\end{equation}
Finally, since $Y(t)$ is a Gaussian variable for all time $t$ we can obtain
the value of the diffusion entropy $S(t)$ using Eq.~(\ref{entropy2}). We see
from Fig.~\ref{figure7} that the numerical diffusion entropy (triangles)
departs from the theoretical expectations at early times ($t$$\ll $$%
1/\lambda $). In fact the initial slope of the numerical diffusion entropy
is $\sim $1 ($0.96$ for $t<0.04s$) instead of 1.5 as expected from Eq.~(\ref
{varY1}). This is due to the overlapping window procedure used by DE \cite{overlaperror}. 
For times $t>>1/\lambda $ the numerical diffusion entropy
reproduces the expected results (slope 0.51 for $0.4s<t<1s$): the
discrepancy for times $t>3s$ is due to the finiteness of the record used.

The DFA curves seems to reproduce both the early time scaling and the later
time scaling: the initial slope is 1.4 ($0.01s<t<0.1s$), while for 2s$<$$t$$<
$10s the DFA curves have slope 0.52. The major discrepancy between the
numerical DFA curves and the expected behavior for $log_{2}F(t)$ (aside from
the shift in values) is that the scaling regime with parameter $\sim $0.5 is
delayed one decade in time ($\sim 2s$ instead of $\sim 0.4s$). This effect
also occurred for in the case of the increments of the variable $X(t)$ of
Eq.~(\ref{langevin1}) (Fig.~\ref{figure5}). 
\begin{figure}[tbp]
\includegraphics[angle=-90,width=1.0\linewidth]{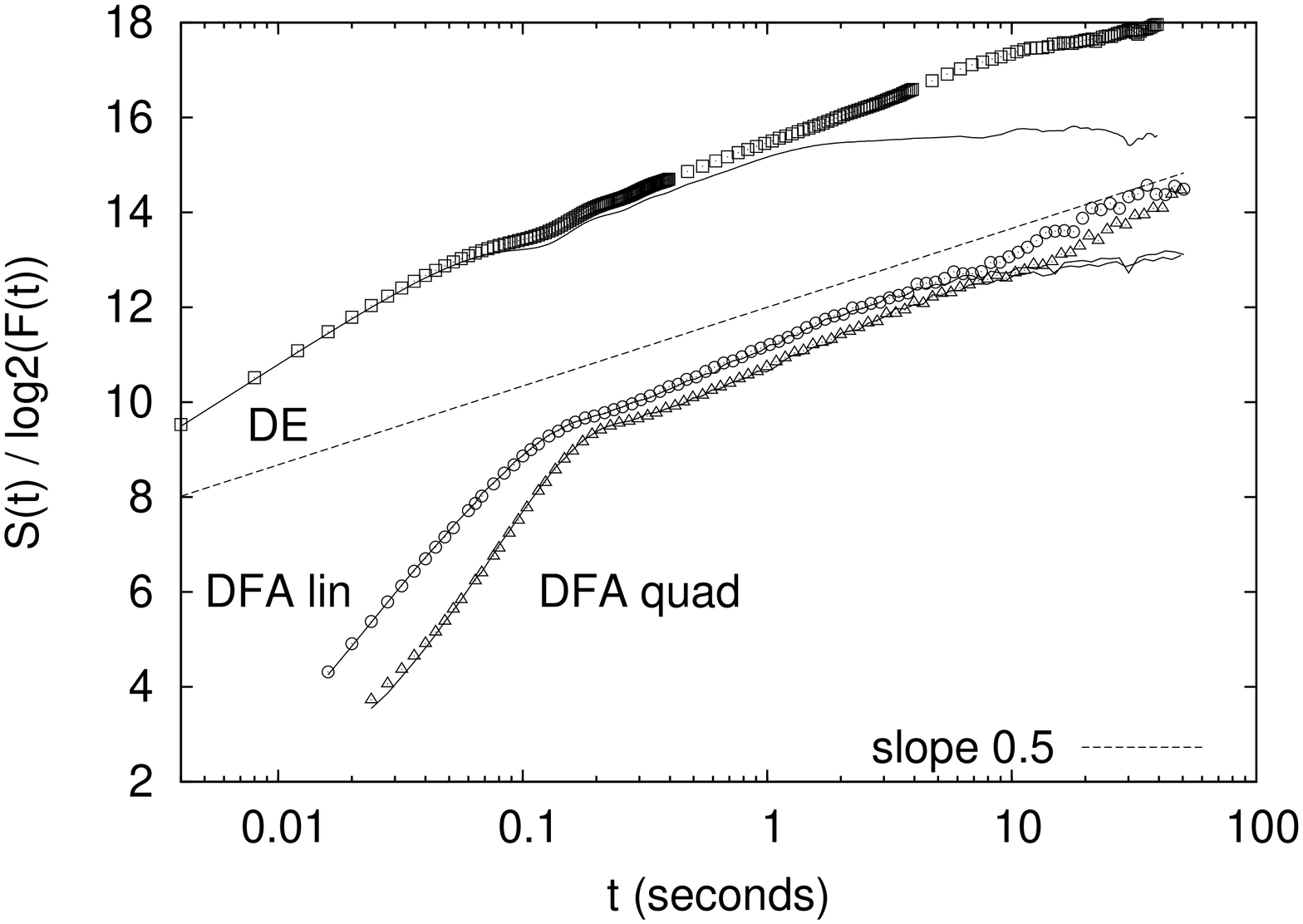}
\caption{DE method and DFA for a EEG record (channel O1) and its best
approximation via the model of Eq.~(\ref{sinusoidal}) (Section \ref{EEGlang}%
). Solid lines indicate the results of DE and DFA relative to the EEG
record. Trianlges indicate the results of DE calculation relative to the
Ornstein-Uhlenbeck process of Eq.~(\ref{sinusoidal}) approximating the EEG
record, while circles (linear detrending) and squares (quadratic deternding)
indicate those of DFA. Finally, the dashed line indicates a logaritmic
increase with a slope of 0.5.}
\label{figure9}
\end{figure}
Fig.~\ref{figure9} shows the results of the DE method and DFA for a EEG
record (channel O1) and its best approximation via the model of Eq.~(\ref
{sinusoidal}) (Section \ref{EEGlang}). Since the DE algorithm does not have
any detrending procedure the presence of an alpha rhythm produces modulation
observed for $t\lesssim 1$s both for the real EEG (solid line) and the its
approximation (squares). The DE of the EEG and that of the model depart at 
$\sim$$0.8-1$s: the EEG entropy saturates while the model DE continues to
increase. The DFA of the real EEG shows a double ``scaling'' regime for the
DFA ($t\lesssim$$0.1$s and $0.4$s$\lesssim$$t\lesssim$4s). At $t$$\sim$4s a
split, similar to that observed in the DE at $t$$\sim$1s, occurs between the
DFA of the EEG which saturates and that of the model which continues to
increase. The slope (we report the values for the DFA with linear detrending
as the values for the quadratic detrending are similar) of the first scaling
regime is 1.75 for the EEG record and 1.73 for the model approximation. 
The second scaling regime has a slope of 0.65 for the EEG and 0.64 for the
model approximation. Are these genuine scaling regimes? As in the similar
case presented in Fig.~\ref{figure9}, these linear regime are not genuine
scaling regimes but are the result of the EEG dynamics not satisfying the
decomposition of Eq.~(\ref{sigplusnoise}) which compromises the capability
of the DFA to detrend the alpha-wave component. For times $t$$>$4s the DFA
curve relative to the EEG starts to bend so that a linear fit is not
feasible. However the model approximation curves keep increasing. The nature
of the saturation observed for both DE and DFA in Fig.~\ref{figure9} has
been recently explained \cite{val08} as being due to high-pass filtering by
the EEG recording apparatus. Valencia \textit{et al}. \cite{val08} show that
the saturation time for the DFA of the EEG record is simply the inverse of
the cutoff frequency $f_{c}$ of the high-pass filter: $0.3$Hz in the present
case.

\section{Coherence of Alpha Wave}

\label{alphawco} 
\begin{figure}[tbp]
\includegraphics[angle=-90,width=1.0\linewidth]{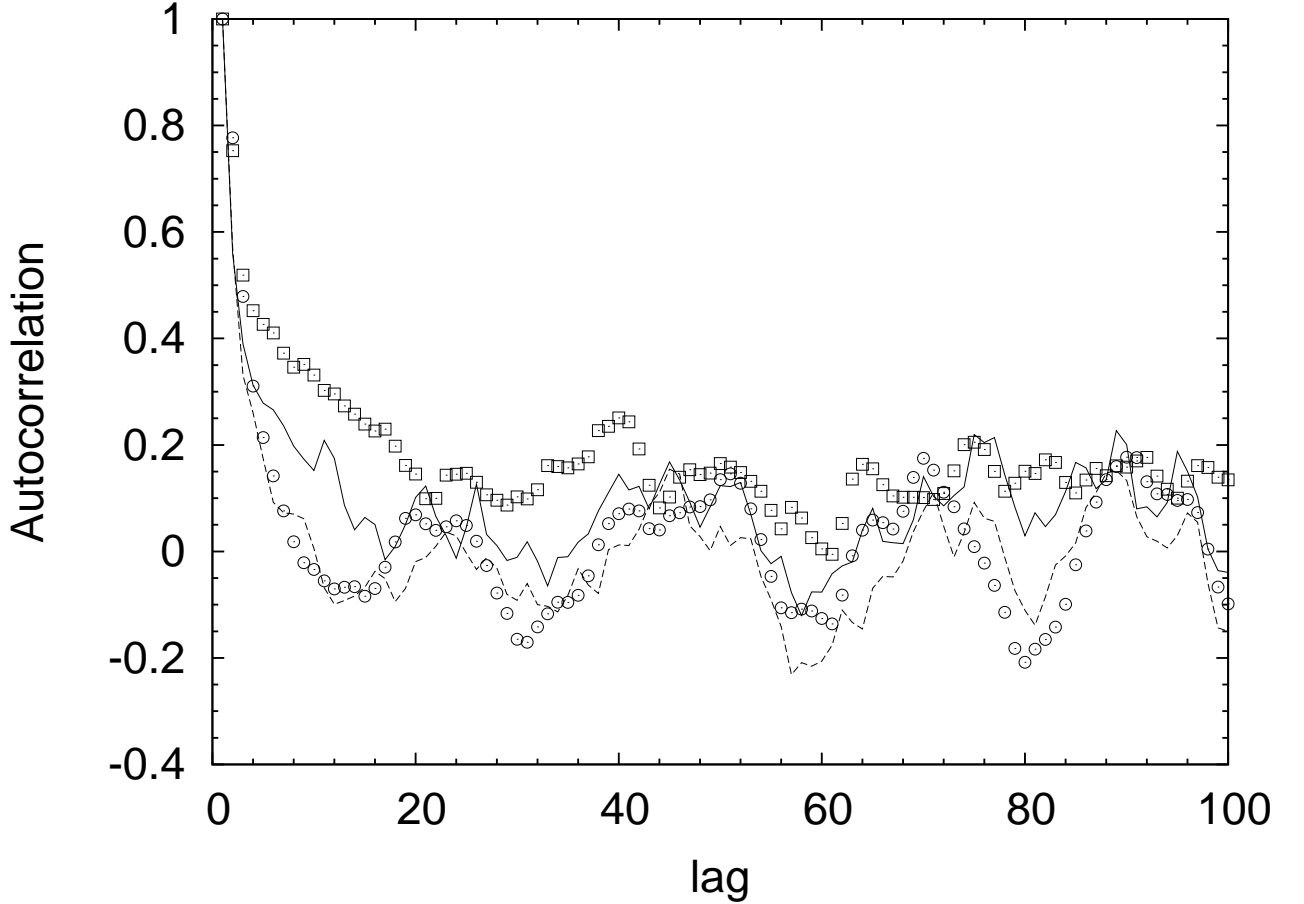}
\caption{Autocorrelation function for the sequence of amplitudes and
frequencies of the alpha rhythm for the channels O1 and C4: empty sqaures
(circles) indicate the autocorrelation function of the sequence $A_{j}$, and solid (dashed) 
lines the autocorrelation function of the sequence $f_{j}$
for the channel O1 (C4).}
\label{figure10}
\end{figure}

\begin{figure}[tbp]
\includegraphics[angle=-90,width=1.0\linewidth]{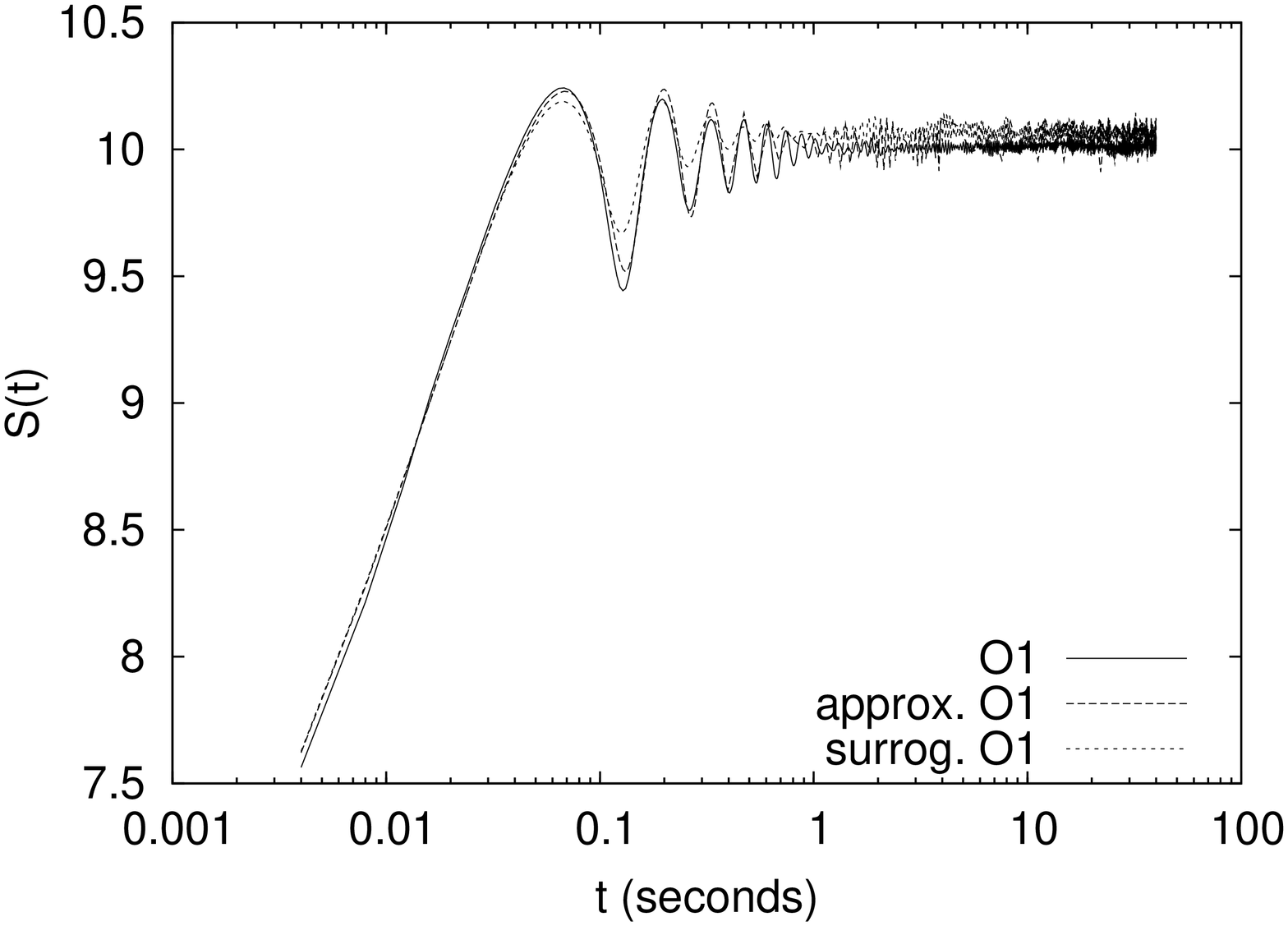}
\caption{Comparison of DE applied to the increments time series: solid line
(EEG channel O1), dashed line (best approximation via Eq.~(\ref{sinusoidal})
and, dot-dashed line (surrogated data obtained applying Eq.~(\ref{sinusoidal}%
) to a shuffled sequence of couples $\{$$A_{j}$,$f_{j}$$\}$).}
\label{figure11}
\end{figure}

Nikulin and Brismar \cite{nikulin05} use the Hilbert
transform of EEG time series to define a sequence of EEG amplitudes, which
is then filtered to obtain a sequence of alpha rhythm amplitudes. DFA is
then applied to this latter sequence to find scaling in the $5-50s$ range.
The scaling parameters change within each channel of a single subject and
among subjects: 0.71 median with a quartile range $0.63-0.81$.
Linkenkaer-Hansen et al. \cite{hansen01} apply the DFA to the sequence of
the moduli of the wavelet transform of the EEG signal in the scale range
corresponding to the alpha rhythm ($8.3-11.7Hz$). They report a scaling
parameter of $0.68\pm 0.07$ (average over the subjects and channels) in the $%
5-300s$ time range. In Fig.~\ref{figure10} we report the autocorrelation
function for the times series of the amplitudes $A_{j}$ and frequency $f_{j}$
for the alpha rhythm for two different channels: O1 and C3. Each couple $\{$$%
A_{j}$,$f_{j}$$\}$ represents the amplitude and frequency of the alpha
rhythm during 0.5s of EEG activity (see Section IIC). Fig.~\ref{figure10}
shows fast decay for the autocorrelation of both the amplitudes and the
frequencies of each channel. The autocorrelation drops to $\sim 0.1-0.2$ at
a lag of $\sim 10$ (which correspond to a coherence time of $\sim$10s) 
after which the autocorrelation function oscillates in
the range $\left( -0.2,0.2\right) $. To quantify the strength of the
correlation between the sequences $A_{j}$ and $f_{j}$ we shuffle the two
time series and use Eq.~(\ref{sinusoidal}) to create a surrogate record. In
Fig.~\ref{figure11} we compare the DE for the EEG increments (channel O1),
with its best approximation via Eq.~(\ref{sinusoidal}) and the surrogate
data obtained applying Eq.~(\ref{sinusoidal}) to a shuffled sequence of
couples $\{$$A_{j}$,$f_{j}$$\}$. We see how shuffling the couples $\{$$A_{j}$%
,$f_{j}$$\}$ results in a sharper attenuation of the periodic alpha
modulation, but does not change the qualitative results. 

\section{Discussion and Conclusions}

The first notable property of the OU Langevin model of EEG time series is
that the resulting EEG diffusive entropy reaches a saturation level. The EEG
entropy saturation indicates that the EEG time series asymptotically carries
a maximum amount of information. Robinson \cite{robinson03} observed this
saturation in the calculation of second moments of EEG time series and
interpreted it as being due to dendritic filtering. The EEG entropy does not
grow indefinitely as would a random process with long-time correlation;
consequently, the EEG time series do not simply scale as had been previously
assumed by a number of investigators \cite{hwa02,lee02,yuan06}. 

Schl\"{o}gl et al. \cite{schogl99} remarked that biosignals typically
saturate due to the limited dynamic range of amplifiers and observed the
saturation of EEG entropy during sleep using data from eight European
laboratories. However we find that the time constant for saturation due to
the limited dynamic range of the amplifier is significantly larger than that
due to physiologic processes in the brain as indicated by the lack of
saturation in Fig. 8. How does this filtering procedure affects the observed
results for the DE and DFA of the EEG increments (Fig.~\ref{figure6})? Is the
saturation effect observed for the DE of the EEG increments genuine or is it
an artefact of high pass filtering? Table I reports the typical values
observed for the dissipation parameter $\lambda $ of the Eq.~(\ref
{sinusoidal}) used to fit the observed DE curve relative to EEG increments.
The time $1$/$\lambda $ is the ``saturation'' time of the DE curve as shown
by Eq.~(\ref{langevin5}). The time $1$/$\lambda $ can be considered as the
``saturation'' time even when an alpha-wave component in present in the EEG
record, since this results in a periodic modulation to Eq.~(\ref{langevin5}%
). Recalling that our EEG records are sampled at 250 Hz, and that from table
I the value $0.04Hz$ can be considered as a typical value for the parameter $%
\lambda $, we obtain a typical saturation time of the order of $0.1s$ which
is considerably smaller than the saturation time $1/\left( 0.3Hz\right)
\simeq 3.3s$ due to the high-pass filtering.

The second notable property of the OU Langevin model is related to the first
and is the dissipation, or negative feedback, produced locally within the
channel of interest. The fluctuation-dissipation relation of Einstein
quantifies the maximum level of the entropy in a closed physical network,
and is given by the ratio of the strength of the additive fluctuations to
the dissipation rate. In the more general OU driven Langevin equation given
here we do not expect the saturation level to be given by this ratio alone,
but to depend on the asymptotic value of the 'variance' $\sigma \left(
t\rightarrow \infty \right) $. Note that the asymptotic 'variance' may not
be independent of time, but contains residual information in the form of low
amplitude beats because of its dependence on the random near-periodic
driver. This mechanism also explains the saturation observed earlier \cite{robinson03} 
by associating the negative feedback with the dendritic
filtering of the signal.

The third notable feature of the OU Langevin model is the attenuated
oscillation of the entropy in time. We reasonably interpret the attenuation
of the modulation of the EEG entropy to be a consequence of the alpha rhythm
not being generated at a single source, but to be a collective property of
the brain being generated at a number of different locations \cite{basar97,patel99}. 
Here the influence of the distributed sources is modeled by
wave packets that persist for a stability time $t_{s}$; one packet is
replaced by another with a slightly different carrier frequency and
amplitude chosen from the empirical spectrogram over time intervals of
length $t_{s}$. The concatenation of these wave packets with fluctuating
frequencies and amplitudes produces a decoherence that attenuates the
modulation of the resulting EEG entropy in time. Both Figures 10 and 11
indicate that the attenuation of the alpha rhythm is dependent on the
statistics of the amplitude and frequency fluctuations and not on their
statistics.

The presence of alpha-rhythm modulation masks \cite{bimbini} any early-time
scaling property of the EEG dynamics. Eq.~(\ref{sinusoidal}) is the simplest
form of a fluctuation-dissipation process that implies the presence of
internal feedback to prevent the occurrence of large excursions of the
electric potential inside the brain. The presence of this negative feedback
mechanism casts doubt on the possibility of understanding EEG records in
terms of a sum of two independent components, noise and trend (signal),
which is the usual assumption made for the DFA method.

Finally, the analysis presented herein supports the notion that alpha
rhythms \cite{basar97} are not passive states, but contain useful
information within the frequency modulation.

The authors thank the Army Research Office for support of this research. The
code of the programs used for the EEG analysis (DE and spectrogram) is
available at http://www.duke.edu/$\sim $mi8/softwaresubpage/C$+$$+$\_programs.html


\begin{thebibliography}{99}
\bibitem{overlaperror} The discrepancy between the observed behavior of the DEA and the theoretical one is 
due to the overlapping window algorythm, Eq.~(\ref{rndwlk}) used in the DEA.
\end{thebibliography}
\end{document}